\documentclass{aa}
 
\usepackage{graphicx,float}

\def \sax {BeppoSAX}

\def \hcm {\hbox {\ifmmode $ atom cm$^{-2}\else atom cm$^{-2}$\fi}}
\def \arcmin {\hbox{$^\prime$}}

\def \chisq {$\chi ^{2}$}

\def \mdot{\dot {\rm M}}
\def \msun{{\rm ~M_{\odot}}}

\def\approxgt{\mathrel{\hbox{\rlap{\lower.55ex \hbox {$\sim$}}
        \kern-.3em \raise.4ex \hbox{$>$}}}}
\def\approxlt{\mathrel{\hbox{\rlap{\lower.55ex \hbox {$\sim$}}
        \kern-.3em \raise.4ex \hbox{$<$}}}}
\newcommand{\mc}{\multicolumn}

\begin{document}

\thesaurus{06.(02.01.2; 08.09.2; 08.14.1; 10.07.3; 13.25.3)}

\title{BeppoSAX study of the broad-band properties of luminous
globular cluster X--ray sources}

\author{L. Sidoli\inst{1}
        \and A.N. Parmar\inst{1}
        \and T. Oosterbroek\inst{1} 
        \and L. Stella\inst{2}
        \and F. Verbunt\inst{3}
        \and N. Masetti\inst{4}
        \and D. Dal Fiume\inst{4}
}
\offprints{L.Sidoli (lsidoli@astro.estec. esa.nl)}

\institute{
       Astrophysics Division, Space Science Department of ESA, ESTEC,
       Postbus 299, NL-2200 AG Noordwijk, The Netherlands
\and  
       Osservatorio Astronomico di Roma, Via Frascati 33,
       Monteporzio Catone, I-00040 Roma, Italy
\and     
       Astronomical Institute, Utrecht University,
       P.O.Box 80000, NL-3508 TA Utrecht, The Netherlands 
\and
       Istituto Tecnologie e Studio Radiazioni Extraterrestri, CNR, 
       Via Gobetti 101, 40129 Bologna, Italy
}
\date{Received 16 October 2000; Accepted 12 December 2000 }

\maketitle
\markboth{Globular cluster X--ray sources}{Globular cluster X--ray sources}

\begin{abstract}

We have performed a detailed study of the broadband spectra of 
the luminous 
($\approxgt 10^{36}$~erg~s$^{-1}$) globular cluster
X--ray sources  using BeppoSAX. With the
exception of X\thinspace2127+119, located in NGC\thinspace7078, all the
other spectra are well represented by a two component model consisting
of a disk-blackbody and Comptonized emission. The measured low-energy
absorptions are in good agreement with those predicted from optical
measurements of the host globular clusters. This implies that there is
little intrinsic X--ray absorption within the binaries themselves, and that
the above spectral model provides a good representation of the low-energy
continua.
The sources can be divided into two groups. 
In the first group, composed of 
3 ultra-compact    (orbital period $<$1~hr) sources,  the disk-blackbody 
temperatures and inner-radii appear physically realistic 
and the Comptonization seed photons temperatures and radii 
of the emission areas 
are consistent with the disk temperatures 
and inner radii.
For all the other sources, the disk-blackbody parameters appear not to be
physically realistic and the Comptonization parameters are unrelated
to those of the disk-blackbody emission.
If this is a spectral signature of ultra-compact  binaries, 
this implies that no other ultra-compact   binaries are 
present among those studied here.
It is unclear why this difference between the two types of binaries 
should exist. 
One possibility may be related to the mass ratio, which is similar 
in the ultracompact systems and binaries containing black holes. 
In the latter systems the soft components are also well-fit with disk-blackbody
models, which appear to have physically realistic parameters.


\end{abstract}

\section{Introduction}
\label{sect:intro}
 
The nature of the X--ray sources in globular clusters has provoked
great observational and theoretical interest.
Surveys of the 
dozen brightest $(\approxgt$$10^{36}$~erg~s$^{-1}$) globular cluster 
X--ray sources
have shown that they share many similarities with the low-mass
X--ray binaries (LMXBs) 
found in the rest of the Galaxy - for example the
presence of X--ray bursts and a lack of coherent pulsations in their persistent 
emission (Hertz \& Wood \cite{h:85}; Parmar et al. \cite{p:89}; 
Verbunt et al. \cite{v:95}).
The presence of Type I bursts indicates that the luminous systems contain
accreting neutron stars rather than black holes, and they are generally
considered to be similar systems to the galactic X--ray burst sources.
Their  formation  is, however,
about a factor $\sim$200 more efficient inside 
globular clusters than in the galaxy itself.

LMXB X--ray spectra are generally modeled by
combining one or more spectral components. At
luminosities $\sim$$10^{37}$~erg~s$^{-1}$ these often consist
of a modified blackbody 
and some form of power-law like
component thought to result from the Comptonization of
cooler photons. At lower luminosities, the
spectra appear to simplify and are often fit with a power-law with an 
exponential cut-off   at high energies (White et al. \cite{w:88}).

In this {\it paper} we present the results of a
systematic survey of the luminous globular cluster X--ray sources 
undertaken as part of the BeppoSAX Core Program. We 
pay particular attention to the spectral properties of the persistent
emission. A long-term goal of such studies is to understand 
how the spectral shape may be used to derive information on 
physically interesting parameters such as the mass accretion rate,
the neutron star mass, radius and spin-rate, the system inclination and
the viscosity of the material in the accretion disk.
For surveys of LMXBs, the study of those located in globular clusters
provides a number of advantanges: (1) the distances to the
many globular clusters have been independently 
determined through main-sequence 
modeling and other methods (see e.g., Carretta et al. \cite{c:00}).
This allows more accurate estimates of luminosity 
than is typical for galactic LMXBs. 
(2) The mean cluster abundances have been determined, allowing
any dependence of X--ray properties on abundance to be examined. (3)
The cluster reddenings have been measured so providing
an independent estimate of low-energy absorption.
For a number of
clusters the reddening is significantly less than typical for galactic LMXBs.  

Results of the BeppoSAX observations of Terzan\thinspace1,
Terzan\thinspace2, NGC\thinspace6441, and NGC\thinspace7078
are to be found in Guainazzi et al. (\cite{g:99}, \cite{g:98}),
Parmar et al. (\cite{p:99}) and Sidoli et al. (\cite{s:00}), respectively.
In addition, NGC\thinspace6440, Terzan\thinspace6, and 
Liller\thinspace1 were observed by BeppoSAX in other programs. 
Results are reported in 
in't   Zand et al. (\cite{i:99}, \cite{i:00}) and Masetti et al.
(\cite{m:00}), respectively.
For these sources   data were  extracted from the  SSD BeppoSAX 
archive and reprocessed using the SAX Data Analysis System (SAXDAS). 

\section{Globular Cluster Properties}
\label{sect:globs}

Table~\ref{tab:log} includes informations on the globular clusters
which are known to contain luminous X--ray sources and have been analysed here. 
Most of the cluster distances, metallicities and extinctions
are taken from Djorgovski (\cite{d:93}). In the case of 
Terzan\thinspace1 more recent estimates of the distance and
extinction obtained using HST are used (Ortolani et al. \cite{o:99}).
For the highly absorbed globular cluster Terzan\thinspace6 the
results of Barbuy et al. (\cite{ba:97}) are used, while the metallicity
of Liller\thinspace1 is taken from the infrared measurements of 
Frogal et al. (\cite{f:95}).

The metallicities of the studied globular clusters range from strongly
underabundant compared to solar (NGC\thinspace7078, [Fe/H] = $-2.17$),
to around solar (Liller\thinspace1, [Fe/H] = +0.25). Such a wide range
of metallicity should allow any dependence of the X--ray properties
on this parameter to be probed. However, we caution that the material being 
accreted does not necessarily have the same metallicity as the host globular
cluster since the companion star may have undergone a non-standard 
evolution during which its envelope composition was altered. 

The extinctions to the studied globular clusters range from low
(${\rm A_v = 0.1}$, NGC\thinspace1851) to strongly absorbed 
(${\rm A_v \sim 10}$, Liller\thinspace1). The amount of extinction  
strongly affects the low-energy cut-off   of the X--ray data and so the
sensitivity to any low-energy spectral components. As an example, 
the X--ray source in NGC\thinspace1851 is visible using BeppoSAX
down to 0.1~keV, while that in Liller\thinspace1 is only detected
above 1~keV.

A summary of all previous observations of the  
Globular Clusters luminous 
X--ray sources
can be found in Sidoli et al. (\cite{s:00b}).

\section{Observations}
\label{sect:obs}

Results from the Low-Energy Concentrator Spectrometer (LECS;
0.1--10~keV; Parmar et al. \cite{p:97}),   Medium-Energy Concentrator
Spectrometer (MECS; 1.8--10~keV; Boella et al. \cite{b:97}),
  High Pressure Gas Scintillation Proportional Counter
(HPGSPC; 5--120~keV; Manzo et al. \cite{m:97}) and   Phoswich
Detection System (PDS; 15--300~keV; Frontera et al. \cite{f:97}) on-board
BeppoSAX 
are presented. All these instruments are coaligned and collectively referred
to as the Narrow Field Instruments (NFI).
The MECS and the LECS are grazing incidence 
telescopes with imaging gas scintillation proportional counters in
their focal planes. 
The non-imaging HPGSPC consists of a single unit with a collimator
that was alternatively rocked on- and 180\arcmin\ off-source 
every 96~s during most of the observations. During the observations
in the year 2000 the collimator remained on-source.
The 
PDS consists of four independent non-imaging 
units arranged in pairs each having a
separate collimator. Each collimator was alternatively
rocked on- and 210\arcmin\ off-source every 96~s during 
the observations.

\begin{table*}
\caption[]{BeppoSAX observations of luminous globular cluster X--ray
sources. T is the MECS exposure and R the energy range used for
spectral fitting.   LMHP indicate whether data from
the LECS, MECS, HPGSPC, and PDS respectively, were included in the fits}
\begin{flushleft}
\begin{tabular}{llrlllrlrrr}
\hline\noalign{\smallskip}
Globular & X--ray &${\rm P_{orb}}$ & \mc{2}{c}{BeppoSAX Observation} & \hfil T 
& \mc{1}{c}{R}  & Inst.
& \mc{1}{c}{d} &Cluster & ${\rm A_v}$\\
Cluster  & Source & (hr) & \hfil Start & \hfil Stop & (ks) & (keV) & 
   & (kpc) &  $[$Fe/H$]$ \\ 
\noalign{\smallskip\hrule\smallskip}
NGC\thinspace1851 & X\thinspace0512$-$401 & $<$0.85     & 2000 Feb 23 12:44 & Feb 25 07:44
&72.2    & 0.1--25  &LM{$-$}P &12.2 & \llap{$-$}1.29 & 0.1\\
Terzan\thinspace2 & X\thinspace1724$-$304 & & 1996 Aug 17 04:29 & Aug 18 05:05
&37.1 & 0.3--100  &LMHP &10.0 & \llap{$-$}0.25 & 4.0\\
Liller\thinspace1 & X\thinspace1730$-$335 & & 1998 Feb 27 03:30 & Feb 27 19:22
&16.2 &  1.0--10  &LM{$-$}{$-$} &7.9 & +0.25 & $\sim$10 \\
                  &                       & & 1998 Mar 02 19:57 & Mar 03 12:35
&29.8 & 1.0--10 &LM{$-$}{$-$} & & &  \\
Terzan\thinspace1 & X\thinspace1732$-$304 & & 1999 Apr 03 12:52 & Apr 04 15:29
& 47.1 & 2.0--10 &{$-$}M{$-$}{$-$} &5.2 &  \llap{$-$}0.71 & 6.8 \\
NGC\thinspace6440 & X{\thinspace}1745$-$203 & & 1998 Aug 26 02:11 
& Aug 26 14:36 & 21.6 & 0.4--100 & LMHP &7.0 &  \llap{$-$}0.34 & 3.5 \\ 
NGC\thinspace6441 & X\thinspace1746$-$370 &5.7 & 1999 Apr 04 18:07 
& Apr 05 20:23
& 49.9 & 0.25--30 &LMHP & 10.7 & \llap{$-$}0.53 & 1.3 \\
Terzan\thinspace6 & X\thinspace1747$-$313 &12.4 & 1998 Sep 06 06:16 
& Sep 06 19:02 
& 28.9 & 0.5--50 &LMHP & 7.0 & \llap{$-$}0.61 & 7.4 \\
NGC\thinspace6624 & X\thinspace1820$-$303 &0.2& 1997 Oct 02 06:45 
& Oct 02 19:90
& 22.3 & 0.3--30 &LMHP & 8.1 & \llap{$-$}0.37 & 0.9 \\
                  &                       & & 1998 Apr 17 04:14 & Apr 18 03:12
& 37.5 & 0.3--30 &LMHP &  \\
                  &                       & & 1998 Sep 23 12:44 & Sep 24 15:30
& 48.3 &  0.3--30 &LMHP  &  \\
NGC\thinspace6712 & X\thinspace1850$-$087 &0.3 & 1997 Apr 24 11:47 
& Apr 25 11:09
& 42.0 & 0.3--50 &LMHP &6.8  & \llap{$-$}1.01 & 1.4 \\
NGC\thinspace7078 & X\thinspace2127+119 &17.1& 1999 Nov 16 01:15 
& Nov 17 00:50 
& 35.5 & 0.1--100 &LM{$-$}P & 10.5 & \llap{$-$}2.17 & 0.2 \\
\noalign{\smallskip\hrule\smallskip}
\end{tabular}
\end{flushleft}
\label{tab:log}
\end{table*}


Table~\ref{tab:log} lists the BeppoSAX observations used in this
survey together with some basic  information about each globular
 cluster  (see Sect.~\ref{sect:globs}). 
The observations of Terzan\thinspace2 and NGC\thinspace6712 were
made prior to the failure of one of the MECS units.
Data from the entire observations were accumulated, except
for the following sources: (1) X\thinspace1730$-$335 (the Rapid Burster, hereafter RB).
BeppoSAX carried out a total of 7 observations, partly analysed by 
Masetti et al. (\cite{m:00}),
 which were extracted from
the archive and examined for the presence of bursts. 
Only two
observations included sufficiently long    quiescent 
intervals for this study (see Table~\ref{tab:log}). 
(2) For the 5.7 hour dip source X\thinspace1746$-$371 the dipping and bursting
intervals discussed in Parmar et al. (\cite{p:99}) were excluded.   
(3) The eclipse observed from the 12.4~hr eclipsing system 
X\thinspace1747$-$313 (In 't Zand et al. \cite{i:00}) was
excluded. 
(4) The bursts seen from X\thinspace0512$-$401, 
X\thinspace1724$-$304, and X\thinspace1745$-$203 were 
excluded. 
Table~\ref{tab:log} also indicates the instruments that were used in the spectral analysis 
of each source.
In the case of the faintest source studied (Terzan\thinspace1, 
0.014~MECS~counts~s$^{-1}$) only the MECS was used, while for the slightly
brighter Liller\thinspace1 observations the LECS and MECS were used.
If available, all the NFI were used for the brightest sources.

Since previous analyses used different software and calibration
releases, all relevant data were re-extracted.
As usual, good data were selected from intervals when the 
elevation angle
above the Earth's limb was $>$$4^{\circ}$ for the LECS and 
$>$$5^{\circ}$ 
for other instruments   and when the instrument
configurations were nominal. 
This was done by using the SAXDAS 2.0.0 data analysis package.
LECS and MECS data were extracted centered on the source positions 
using radii of 8\arcmin\ and 4\arcmin, respectively, 
corresponding to about 95\% of the total source photons. 
In the case of the RB
extraction radii of 4\arcmin\ and
2\arcmin\ were used in order to minimize the contribution from the 
32\arcmin\ distant bright X--ray source 4U\thinspace1728$-$337. 
Similarly, data from the non-imaging instruments were not included
in the RB fits.
Background subtraction for the imaging instruments
was performed using standard files; this is not a critical procedure for the
bright sources studied here (with the exception of Terzan\thinspace1). 
Background subtraction for the non-imaging instruments was carried out 
using the offset pointing intervals.
In the case of observations were the HPGSPC remained
coaligned, background subtraction was 
performed using data obtained when the instrument
was looking at the dark Earth.

The overall spectra extracted as described above
were first investigated by simultaneously
fitting data from as broad a band as possible. 
The LECS and MECS spectra were rebinned to oversample the full
width half maximum of the energy resolution by
a factor 3 and to have additionally a minimum of 20 counts 
per bin to allow use of   $\chi^2$ statistics. 
The HPGSPC and PDS 
spectra were rebinned using the standard techniques in SAXDAS.
Data were selected from within the energy ranges
0.1--8.0~keV (LECS), 1.8--10~keV (MECS), 8.0--20~keV (HPGSPC),
and 15--100~keV (PDS) whenever a sufficient number of 
counts were obtained (see Table~\ref{tab:log}).
The photo-electric absorption
cross sections of Morrison \& McCammon (\cite{m:83}) 
and the
solar abundances of Anders \& Grevesse (\cite{a:89})
 are used throughout.
Factors were included in the spectral fitting to allow for normalization 
uncertainties between the instruments. 
These factors were constrained
to be within their usual ranges during the fitting (Fiore et al. \cite{f:99}). 
An additional uncertainty
of 1\% was added quadratically to the uncertainties to account for systematic 
uncertainties in instrumental responses.
                                                     
Since we wish to investigate the spectral properties of the sources,
we first attempted to find a general spectral model that could be successfully
applied to all the sources. Many luminous LMXB   are fit with a two component
model consisting of a blackbody, or blackbody-like component which might
represent emission from an optically thick  accretion disk or from the
neutron star surface, together with a  Comptonized component 
which may be interpreted as   emission
from a hot inner disk region or a 
boundary layer between the disk and the neutron star.
For the soft component we tried both a single blackbody model and
a disk-blackbody component, in the formulation of 
Mitsuda et al. (\cite{m:84}) and Makishima et al. (\cite{m:86}).
The disk-blackbody model assumes that the gravitational energy
released by the accreting material is locally dissipated into
blackbody radiation, that the accretion flow is continuous throughout
the disk, and that the effects of electron scattering are negligible.
There are only two parameters in this  model,
${\rm r_{in}({\cos}i)^{0.5}}$ where ${\rm r_{in}}$ is the innermost
radius of the disk, {\em i} is the inclination angle of the disk and
${\rm kT_{in}}$ the blackbody effective temperature at ${\rm r_{in}}$.    
Some limitations of this model are discussed in
Merloni et al. (\cite{mfr:99}).

For the Comptonized  component the {\sc xspec} model
{\sc comptt} (Titarchuk \cite{ti:94};
Hua \& Titarchuk \cite{h:95}; Titarchuk \& Lyubarskij \cite{tl:95}),
which self-consistently calculates the spectrum produced
by the Comptonization of soft photons in a hot plasma, 
 was used.
This model
contains as free parameters the temperature of the Comptonizing
electrons, kT${\rm _e}$, the plasma optical depth with respect to
electron scattering
${\rm \tau_p}$ and the input
temperature of the soft photon (Wien) distribution (kT${\rm _0}$). A
spherical geometry was assumed for the comptonizing region.

Fit result to the broad-band spectra with the single blackbody plus
Comptonization and disk-blackbody plus Comptonization are reported 
in Tables~\ref{tab:spec2} and~\ref{tab:spec}
respectively. 
A  single temperature blackbody  plus a cut-off   power-law model
was also tried, but with  unacceptable results in the majority of the 
sources.
Since   the disk-blackbody plus Comptonization  emission (Table~\ref{tab:spec}) 
always gives a better  (or equally good)
fit than  single blackbody plus Comptonization emission (Table~\ref{tab:spec2}), 
we will refer
to it   as the ``standard'' model.

\begin{table*}
\caption[]{BeppoSAX spectral results using the   {\sc comptt} and
single-blackbody model. For globular clusters with multiple observations,
the results are given in the same order as in Table~\ref{tab:log}.
L is the 0.1--100 keV (assumed isotropic, in units of 10$^{35}$~erg~s$^{-1}$) 
source luminosity using the distances given in 
Table~\ref{tab:log} and assuming that
the measured ${\rm N_H}$ is interstellar. The units of ${\rm N_H}$ are
$10^{22}$~atom~cm$^{-2}$. f$_{\rm bb}$ is the absorption corrected 0.1--100~keV 
ratio of  blackbody to {\sc comptt} fluxes. 
All uncertainties are quoted at 90\%
confidence}
\begin{flushleft}
\begin{tabular}{llllllllrr}
\hline\noalign{\smallskip}
Globular & \mc{1}{c}{L} &  \mc{1}{c}{${\rm N_H}$} 
&\mc{1}{c}{k${\rm T_0}$} & \mc{1}{c}{kT${\rm _e}$} 
& \mc{1}{c}{${\rm \tau_p}$} & \mc{1}{c}{${\rm kT_{bb}}$} 
& \mc{1}{c}{${\rm R_{bb}}$} & \mc{1}{c}{f$_{\rm bb}$}  & $\chi^2$/dof \\
Cluster  &  & & \mc{1}{c}{(keV)} & \mc{1}{c}{(keV)} & &\mc{1}{c}{(keV)} & \mc{1}{c}{(km)} & & \\
\noalign{\smallskip\hrule\smallskip}
NGC\thinspace1851 & $43$ 
& $0.003 \pm ^{0.005} _{0.003}$ &
$0.20 \pm ^{0.02} _{0.03} $ & $2.4 \pm ^{0.2} _{0.1}$ & $17.4 \pm ^{1.0}
_{1.3}$ & $0.56 \pm ^{0.04} _{0.02}$ & $6.7 \pm ^{0.86} _{1.4}$ &0.14 & 153.2/97 \\
Terzan\thinspace2 & $200$  
& $ 0.83 \pm 0.06$ & $1.11 \pm ^{0.15} _{0.03}$ 
& $36 \pm ^{110} _{6}$ & $2.8 \pm ^{0.4} _{2.4}$ & $0.66 \pm ^{0.07} _{0.06}$ 
& $10.7 \pm ^{0.35} _{0.80}$ & 0.15 & 197.1/178 \\
Liller\thinspace1 &  $8.7$ & 
$1.4 \pm ^{1.0} _{0.4}$ & $<$4.2  
  & $2.5 \pm ^{2.3} _{0.7}$ & $17 \pm ^{23} _{10}$ 
& $<$1.5 & $<$4.0 & $<$0.14 & 100.7/90 \\
   &  $2.5  $ &   
$1.3 \pm ^{0.9} _{1.1}$ & $0.43 \pm ^{0.22} _{0.40}$ 
& $9.1 \pm ^{100} _{6.8}$ & $ 7.3 \pm ^{7.7} _{4.3}$ & \dots & \dots & \dots 
& 71.0/57\\
Terzan\thinspace1&  $0.059  $ & $<$0.66 &
$<$0.48 & $>$2.2 & $<$20 & \dots & \dots & \dots & 18.4/11\\
NGC\thinspace6440 & $66$ &  
$0.38 \pm 0.06$ & $0.50 \pm ^{0.03} _{0.04} $ & $9.6 \pm 0.1$ & 
$8.6 \pm ^{0.70} _{0.80} $ & $1.3 \pm ^{0.3} _{0.1}$ 
& $0.87 \pm ^{0.26} _{0.33}$ 
& 0.05 & 101.5/107 \\
NGC\thinspace6441 & $140 $
& $0.20 \pm ^{0.25} _{0.06} $ & $0.51 \pm ^{0.04} 
_{0.07}$ & $2.35 \pm ^{0.09} _{0.04}$ & $ 17.2 \pm ^{0.30} _{0.75}$ & 
$0.27 \pm  ^{0.27} _{0.08}$ & $28 \pm ^{61} _{11}$ & 0.04 & 117.6/106\\
Terzan\thinspace6 & $59$ & $1.3 \pm ^{0.01} _{0.02}$ 
& $1.34 \pm ^{0.21} _{0.02}$ &
$16.0 \pm ^{0.54} _{1.0}$ & $2.0 \pm ^{0.20} _{0.1}$ 
& $ 0.65 \pm {0.05}$ & $6.7 \pm ^{1.0} _{1.4}$ 
& 0.29 & 144.5/119 \\
NGC\thinspace6624 & $420  $ & $0.10 \pm ^{0.11} _{0.03}$ &
$0.38 \pm 0.02$ & $2.97 \pm ^{0.03} _{0.08}$ & $13.3 \pm 0.2$ 
& $0.21 \pm ^{0.01} _{0.05}$ & $88 \pm 2$ 
& 0.05 & 268.2/120 \\
                  & $500 $ & $<$$0.015$  & 
$0.38 \pm ^{0.03} _{3.0} $  &$3.0 \pm ^{0.01} _{0.03}$ 
&   $13.3 \pm{0.1}$ & 
$1.5 \pm ^{0.01} _{0.06}$ & $1.9\pm ^{0.07} _{0.01}$ 
 &  0.05   &    232.5/117 \\
                  & $560 $ &  $0.25 \pm ^{0.02} _{0.01}$   & 
$0.07 \pm ^{0.02} _{0.02}$  &  $2.7 \pm ^{0.15} _{0.01}$  &
$15.8 \pm ^{0.2} _{1.0}$  &   $0.64 \pm ^{0.16} _{0.01}$  &  
$14.4 \pm ^{0.3} _{5.7}$ & $0.09$ &  588.4/134             \\
NGC\thinspace6712 & $13 $  &  
$0.17 \pm ^{0.05} _{0.01} $ & $0.60 \pm ^{0.20} _{0.08}$ 
& $91 \pm ^{10} _{80}$ &
$1.2 \pm ^{0.05} _{0.65}$ & $0.43 \pm ^{0.02} _{0.01}$ 
&$8.9 \pm ^{1.0} _{0.2}$ & 0.34 & 181.1/154 \\
NGC\thinspace7078  &  $90 $ 
& $0.20 \pm ^{0.05} _{0.20}$ & 
$0.042 \pm ^{0.030} _{0.039}$ & $37 \pm ^{30} _{35}$ & $ 2.9 \pm ^{+\infty}
_{0.4}$ & $1.1 \pm 0.05$  
&$3.3 \pm ^{0.80} _{0.30}$ & 0.24 & 375.5/124\\
\noalign{\smallskip\hrule\smallskip}
\end{tabular}
\end{flushleft}
\label{tab:spec2}
\end{table*}

\begin{table*}
\caption[]{BeppoSAX spectral results using the standard {\sc comptt} and
disk-blackbody model. For globular clusters with multiple observations,
the results are given in the same order as in Table~\ref{tab:log}.
L is the 0.1--100 keV (assumed isotropic, in units of 10$^{35}$~erg~s$^{-1}$) 
source luminosity using the distances given in 
Table~\ref{tab:log} and assuming that
the measured ${\rm N_H}$ is interstellar. The units of ${\rm N_H}$ are
$10^{22}$~atom~cm$^{-2}$. f is the absorption corrected 0.1--100~keV 
ratio of disk-blackbody to {\sc comptt} fluxes. 
$\theta$ is the inclination angle of the disk.
All uncertainties are quoted at 90\%
confidence}
\begin{flushleft}
\begin{tabular}{llllllllrr}
\hline\noalign{\smallskip}
Globular & \mc{1}{c}{L} &  \mc{1}{c}{${\rm N_H}$} 
&\mc{1}{c}{k${\rm T_0}$} & \mc{1}{c}{kT${\rm _e}$} 
& \mc{1}{c}{${\rm \tau_p}$} & \mc{1}{c}{${\rm kT_{in}}$} 
& \mc{1}{c}{${\rm r_{in}} (\cos \theta)^{0.5}$} & \mc{1}{c}{f}  & $\chi^2$/dof \\
Cluster  & & & \mc{1}{c}{(keV)} & \mc{1}{c}{(keV)} & &\mc{1}{c}{(keV)} & \mc{1}{c}{(km)} & & \\
\noalign{\smallskip\hrule\smallskip}
NGC\thinspace1851 & $47$ 
& $0.026 \pm ^{0.030} _{0.005}$ &
$0.33 \pm ^{0.06} _{0.09} $ & $2.8 \pm ^{0.20} _{0.50}$ & $14.0 \pm ^{0.5}
_{0.9}$ & $0.30 \pm ^{0.07} _{0.21}$ & $17 \pm ^{99} _{2.8}$ &0.14 & 126.1/97 \\
Terzan\thinspace2 & $220 $  
& $ 0.78 \pm 0.05$ & $0.63 \pm 0.02$ 
& $27 \pm ^{10} _{5}$ & $3.8 \pm 0.7$ & $ 3.3 \pm 0.3$ 
& $0.28 \pm {0.05}$ & 0.12 & 191.5/178\\
Liller\thinspace1 &  $9.6 $ & 
$2.8 \pm ^{0.5} _{1.6}$ & $0.44 \pm ^{0.24} _{0.44}$ 
  & $3.0 \pm ^{1.3} _{0.5}$ & $6.08 \pm ^{0.42} _{0.25}$ 
& $<$0.50 & $<$37,000 & $<$0.05 & 100.6/90\\
   &  $2.5  $ &   
$1.3 \pm ^{0.9} _{1.1}$ & $0.43 \pm ^{0.22} _{0.40}$ 
& $9.1 \pm ^{100} _{6.8}$ & $ 7.3 \pm ^{7.7} _{4.3}$ & \dots & \dots & \dots 
& 71.0/57\\
Terzan\thinspace1&  $0.059  $ & $<$0.66 &
$<$0.48 & $>$2.2 & $<$20 & \dots & \dots & \dots & 18.4/11\\
NGC\thinspace6440 & $71 $ &  
$0.47 \pm 0.07$ & $0.50 \pm 0.04 $ & $8.7 \pm ^{1.3} _{0.8}$ & 
$9.9 \pm ^{0.8} _{1.4} $ & $1.98 \pm ^{0.33} _{0.13}$ 
& $0.47 \pm ^{0.15} _{0.20}$ 
&0.14 & 84.8/107\\
NGC\thinspace6441 & $140$
& $0.26 \pm 0.02 $ & $0.54 \pm ^{0.04} 
_{0.06}$ & $25 \pm ^{225} _{24.5}$ & $ 0.40 \pm ^{1.6} _{0.40}$ & 
$2.82 \pm 0.05$ & $0.92 \pm ^{0.03} _{0.04}$ & 10 & 107.3/106\\
Terzan\thinspace6 & $67$ 
& $1.39 \pm 0.08$ &
$0.66 \pm ^{0.06} _{0.10}$ & $8.3 \pm ^{1.7} _{1.3}$ 
& $ 6.3 \pm ^{2.7} _{0.4}$ & 
$2.65 \pm ^{0.15} _{0.06}$ & $0.60 \pm ^{0.09} _{0.03}$ 
& 2.16 & 136.5/119 \\
NGC\thinspace6624 & $480 $ & $0.16 \pm ^{0.003} _{0.003}$ &
$0.574 \pm ^{0.004} _{0.004} $ & $2.72 ^{0.01} _{0.01}$ & $15.3 \pm ^{0.05} _{0.05}$ & $0.57 \pm ^{0.05} _{0.05}$ 
&$22 \pm{0.13}$ 
& 0.36 & 152.1/120 \\
                  & $570 $ & $0.16 \pm ^{0.07} _{0.007}$  & 
$0.460 \pm ^{0.02} _{0.07}$  &   $2.86 \pm ^{0.07} _{0.01}$ & 
$13.8 \pm 0.1$ & $0.37 \pm ^{0.011} _{0.005}$ 
&$45 \pm ^{40} _{3}$ &  0.18   &     182.2/117 \\
                  & $660 $ &  $0.27 \pm ^{0.01} _{0.04}$   & 
$0.0639 \pm ^{0.020} _{0.040}$  &  $2.77 \pm ^{0.13} _{0.06}$  &
$15.7 \pm ^{1.43} _{0.10}$  &   $1.06 \pm ^{0.47} _{0.021}$  &  
$4.9 \pm ^{0.8} _{1.8}$ & $0.12$ &  627.7/134             \\
NGC\thinspace6712 & $19 $  &  
$0.39 \pm ^{0.01} _{0.03} $ & $0.81 \pm ^{0.11} _{0.19}$ 
& $71 \pm ^{7} _{25}$ &
$1.70 \pm ^{0.15} _{1.00}$ & $0.63 \pm ^{0.05} _{0.02}$ 
&$4.5 \pm ^{0.15} _{0.75}$ & 0.76 & 160.0/154\\
NGC\thinspace7078  &  $65 $ 
& $0.06 \pm ^{0.01} _{0.03}$ & 
$0.036 \pm ^{0.0095} _{0.006}$ & $92 \pm ^{4.6} _{3.4}$ & $ 2.6 \pm ^{0.1}
_{0.2}$ & $1.92 \pm 0.04$  
&$1.1 \pm{0.009}$ & 2.0 & 284.8/124\\
\noalign{\smallskip\hrule\smallskip}
\end{tabular}
\end{flushleft}
\label{tab:spec}
\end{table*}


\begin{figure*}
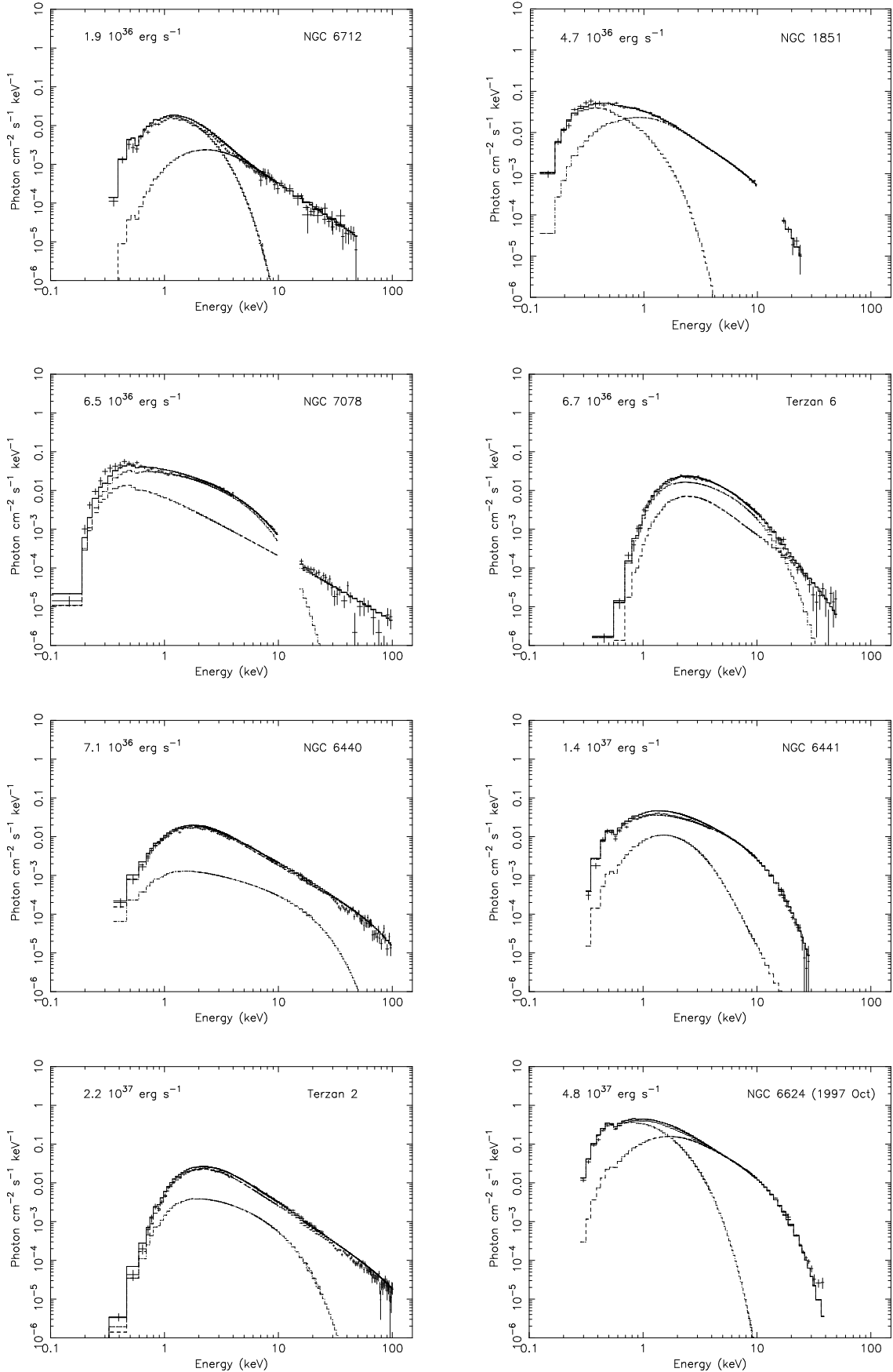

\hbox{\hspace{0.5cm}
\includegraphics[height=6.9cm,angle=-90]{ngc6712_ufs.ps}
\hspace{1.0cm}
\includegraphics[height=6.9cm,angle=-90]{ngc1851_ufs.ps}}
\vbox{\vspace{0.15cm}}

\hbox{\hspace{0.5cm}
\includegraphics[height=6.9cm,angle=-90]{ngc7078_ufs.ps}
\hspace{1.0cm}
\includegraphics[height=6.9cm,angle=-90]{terzan6_ufs.ps}}
\vbox{\vspace{0.15cm}}

\hbox{\hspace{0.5cm}
\includegraphics[height=6.9cm,angle=-90]{ngc6440_ufs.ps}
\hspace{1.0cm}
\includegraphics[height=6.9cm,angle=-90]{ngc6441_ufs.ps}}
\vbox{\vspace{0.15cm}}

\hbox{\hspace{0.5cm}
\includegraphics[height=6.9cm,angle=-90]{terzan2_ufs.ps}
\hspace{1.0cm}
\includegraphics[height=6.9cm,angle=-90]{ngc6624_ufs.ps}}

\caption[]{The BeppoSAX spectra of each of the globular cluster
X--ray sources obtained using the standard model, 
in order of increasing 0.1-100 keV luminosity.
The contributions of the two components
(where evident) are shown separately. Except for
NGC\thinspace6441 the {\sc comptt}
is the one extending to higher energies}
\label{fig:spec}
\end{figure*}

\section{Spectral Fit Results: the Standard Model}
\label{sect:spectrum}

The results of fitting the standard model to all of the extracted spectra
are given in Table~\ref{tab:spec}. 
All spectral uncertainties
and upper-limits are given at 90\% confidence.
The fits to all the spectra are
acceptable with the exception of NGC\thinspace7078 where the $\chi^2$ is
284.8 for 124 degrees of freedom (dof), 
and the second and third observation 
of the source in NGC\thinspace6624. 
Since a different model is  
a better representation of the NGC\thinspace7078 X--ray spectrum, probably
due to
the ADC nature of this source (see Sidoli et al. \cite{s:00}), we  
exclude it from the discussion, though the results of the fitting
procedures have been included in Tables~\ref{tab:spec2} and~\ref{tab:spec},
for completeness.
We note that an acceptable
fit (a $\chi ^2$ of 100.0 for 109 dof) can also be obtained without the 
disk-blackbody component for NGC\thinspace6440. 
In the case of Liller\thinspace1
and Terzan\thinspace1 the sources are too faint to derive accurate
spectral parameters. Moreover Liller~1 was not sampled at  
energies greater than 10 keV due to the reasons explained in Sect.~\ref{sect:obs}. 
We include results from these sources in
Tables~\ref{tab:spec2} and~\ref{tab:spec} for completeness, 
but do not consider them further.
Fig.~\ref{fig:spec} shows the photon spectra of the remaining sources 
derived using the standard model. This figure clearly
illustrates the importance of the broad-band coverage afforded by
BeppoSAX in clearly separating the two components. Bolometric
luminosities have been calculated using the energy range 0.1--100~keV
and assuming that the observed  ${\rm N_H}$ is interstellar.
The results in Sect~\ref{subsect:nh} indicate  that for the majority
of sources this   assumption is valid.

\subsection{Low-energy absorption}
\label{subsect:nh}

We first compared the fitted values of low-energy absorption, ${\rm N_H}$,
with the values predicted using the relation between the optical extinction,
${\rm A_v}$, and absorption of 
${\rm N_H [ cm^{-2}/A_v] = 1.79 \times 10^{21}}$ derived by
Predehl \& Schmitt (\cite{ps:95}). If there is no additional absorption
present in the X--ray binaries themselves, 
and their low-energy X--ray continua have been properly characterized, 
then a good agreement is expected between the
values derived from the globular cluster ${\rm A_v}$ measurements,
and those from the LMXB   X--ray spectra.
Fig.~\ref{fig:nh} shows the measured values of ${\rm N_H}$ taken from
Table~\ref{tab:spec} plotted against the optically 
derived values of ${\rm N_H}$. The straight line shows the locus
of points with equal optical and X--ray derived ${\rm N_H}$ values.

In general, there is good agreement between the two absorption measurements. 
In the case of X\thinspace1745$-$203, located in NGC\thinspace6440
the X--ray derived ${\rm N_H}$ is $\sim$0.75 that expected (a difference
of $1.5 \times 10^{21}$~atom~cm$^{-2}$).
This may suggest the presence
of an additional soft X--ray component that, not being included
in the spectral model, leads to an underestimate of the absorbing column.
Alternatively, the overall X--ray continuum model might be incorrect.
We note that also the blackbody plus {\sc comptt} model does not provide a good 
estimate of the column density, being even much lower.
In 't Zand et al. ({\cite{i:99}) find that the spectrum of
X\thinspace1745$-$203 may also be modeled using either a broken power-law
and blackbody, or a high-energy cut-off power-law, or a bremsstrahlung
and blackbody. 
In these 3 cases, the derived values of
${\rm N_H}$ are in the range 6.7--9.0$\times 10^{21}$~atom~cm$^{-2}$,
similar to that predicted from the ${\rm A_v}$ measurement of
$6.2 \times 10^{21}$~atom~cm$^{-2}$.
In the case of X\thinspace1850--087
located in NGC\thinspace6712 the X--ray derived ${\rm N_H}$, 
of $3.9 \times 10^{21}$~atom~cm$^{-2}$,
 is
significantly higher  
than that derived from the optical of $2.5 \times 10^{21}$~atom~cm$^{-2}$.
This may suggest the presence of extra absorbing material in this system.
However, when fitting this spectrum with a single blackbody plus Comptonization,
a much lower column density is found ($1.7 \times 10^{21}$~atom~cm$^{-2}$).

\begin{figure}
\hbox{\hspace{0.1cm}
\includegraphics[width=8.5cm,angle=0]{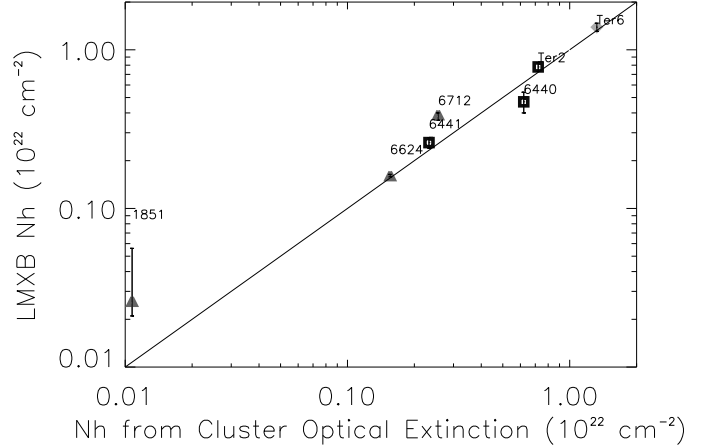}}
  \caption[]{The measured values of ${\rm N_H}$ 
             (Table~\ref{tab:spec}) plotted against the optically 
             derived values of ${\rm N_H}$. The straight line shows 
             the location of points with equal optical and X--ray 
             derived ${\rm N_H}$ values. Here and in all other figures, 
             the light diamonds mark the X--ray binaries known 
             to harbor sub-giant companions (P$_{orb}>10$~hr), the
             triangles the ultra-compact   binaries (P$_{orb}<1$~hr), 
             while the squares indicate all other sources.
             The meaning of the symbols is the same in all the figures }    
   \label{fig:nh}
\end{figure}

\subsection{Disk-blackbody component}
\label{subsect:diskbb}

The soft emission from luminous  globular cluster sources
is  succesfully fit 
with a multicolor disk-blackbody component. 
If we interprete this model ``literally", its
normalization 
translates into 
a lower limit for the innermost radius r$_{\rm in}$ of 
the accretion disk, R$_{\rm in}$=${\rm r_{in}({\cos}i)^{0.5}}$. 
Except for NGC\thinspace1851 and NGC\thinspace6624 
(where the inclination cannot be constrained),
the values observed are generally low,
thus requiring a quite high inclination angle of the binary
systems, for r$_{\rm in}$ to be larger than 
the last stable orbit for an
accretion disk around a neutron star (6GM/c$^{2}$$\sim$9 (${\rm M}/{\rm\msun}$)~km).
We note that still higher inclinations would be derived if the Merloni et al. (\cite{mfr:99})
correction to the disk-blackbody model were taken into account.
For several globular cluster sources, a confirmation 
of their high inclination angles   
comes from X--ray and optical observations. 
In particular, the presence of
strong sinusoidal--like modulations in the X--ray flux  
(as in the case of the
ADC source NGC\thinspace7078), partial X--ray eclipses  
of the central source by the
companion star (Terzan 6), X--ray dips (NGC\thinspace6441) 
are  all indicative of 
inclination angles $>$60--70$^{\circ}$ (Frank et al. \cite{f:87}).
 
The temperature kT$_{{\rm in}}$  at the inner disk radius  is  0.5--3.5~keV, 
with the three ultra-compact   binaries
(NGC\thinspace1851, NGC\thinspace6712, NGC\thinspace6624; P$_{\rm orb}<1$~hr) 
displaying   significantly lower
temperatures ($<$1 keV) compared to all other sources (1.5--3.5 keV). 
The variation of kT$_{{\rm in}}$ and inner radius of the
accretion disk is displayed in Fig.~\ref{fig:ktinrin}, while the variation 
with the disk luminosity, measured in the 0.1--100 keV energy range, 
is shown in Fig.~\ref{fig:rinlum} and Fig.~\ref{fig:summary}.
   
\begin{figure}
\hbox{\hspace{0.0cm} 
\includegraphics[width=8.5cm,angle=0]{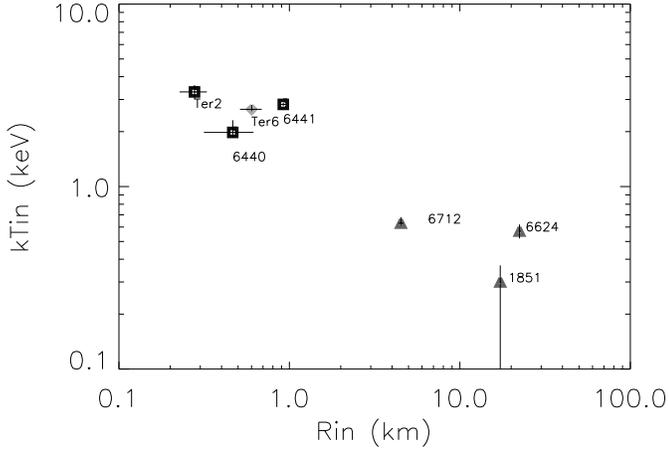}}
  \caption[]{Variation of the disk-blackbody temperature kT$_{{\rm in}}$ 
  	     and the lower limit to the inner radius of the accretion disk   
}
\vbox{\vspace{0.15cm}}
  \label{fig:ktinrin}
\end{figure}


\begin{figure}
\hbox{\hspace{0.0cm} 
\includegraphics[width=8.5cm,angle=0]{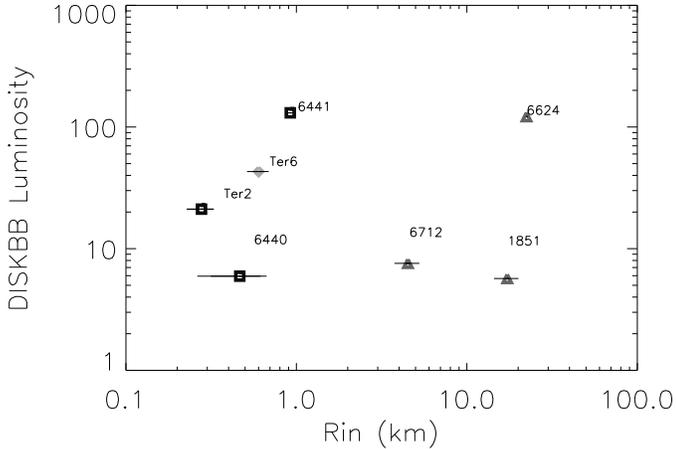}}
  \caption[]{Variation of the lower limit to the inner radius 
of the accretion disk and the luminosity of the disk-blackbody 
component, 
in the 0.1--100 keV band 
(in units of  $10^{35}$~erg~s$^{-1}$) 
}
  \label{fig:rinlum}
\end{figure}

In the disk--blackbody model, the bolometric luminosity, L$_{\rm bol}$, 
of the accretion disk
can be related to the temperature kT$_{\rm in}$ and the mass M 
of the compact object  
(see e.g., Eqs. 9 and 11 in Makishima et al. \cite{mkm:00}) as  
L$_{\rm bol}\propto{\rm M}^{2}\cdot{\rm T_{in}}^{4}$.
A summary of the parameters derived from the disk--blackbody component is 
shown in Fig.~\ref{fig:summary}, where the globular cluster sources
appear to be divided into two groups: the ultra-compact   binaries, 
with kT$_{\rm in}\approxlt$1~keV,
the inferred masses of which are  compatible with that of a neutron star, 
and sources with kT$_{\rm in}\approxgt$1~keV, 
located on
grids with much lower inferred masses for the  compact object. 
Since all these X--ray binaries  (except Terzan~6) 
are known to harbor a neutron star, 
due to their Type I X--ray bursting activity, 
either the multi--colour disk--blackbody is not
an accurate model for the soft part of the spectrum, 
or their luminosity has been
underestimated, possibly due to a very high inclination 
angle of the binary systems.
Another explanation may be that the parameters 
derived in the framework of the disk-blackbody
model cannot be related directly to physical quantities 
(see Merloni et al. \cite{mfr:99}).

\begin{figure}
\hbox{\hspace{-0.5cm} 
\includegraphics[height=6.5cm,angle=0]{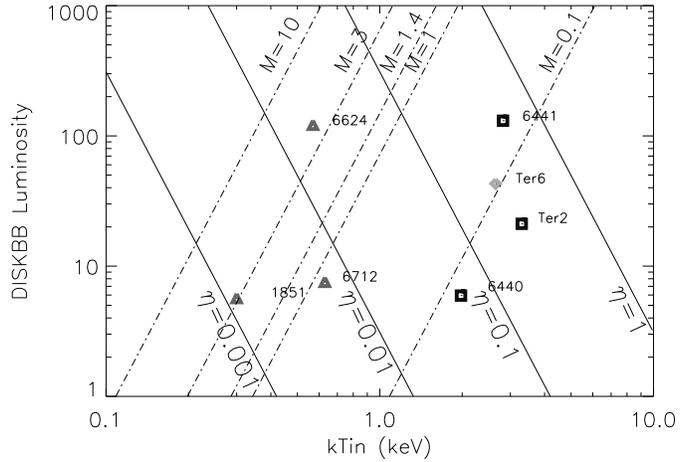}}
  \caption[]{Relation between the disk-blackbody luminosity 
measured in the 0.1--100~keV energy range (in units of  $10^{35}$~erg~s$^{-1}$) and the 
temperature kT$_{\rm in}$ at the inner accretion disk radius.  
Dash--dotted lines represent loci of constant mass for the 
neutron star (in units of solar masses), while 
the solid lines show loci of constant $\eta$, that is the 
luminosity normalized to the Eddington luminosity (assumed 
to be 1.5$\times10^{38}$~erg~s$^{-1}$ for a solar mass compact object) 
}
  \label{fig:summary}
\end{figure}

\subsection{Comptonization component}
\label{subsect:comptt}

Thermal Comptonization  (Titarchuk \cite{ti:94})
provides a good fit to the high energy region
of the globular cluster sources spectra. 
The relations between the different parameters derived from the spectral analysis are shown in 
Figs.~6--11. 
The Comptonizing plasma  
has  temperatures kT$_{\rm e}$ ranging from few keV up to the extremely high value of $\sim$92~keV
in the case of the ADC source in NGC\thinspace7078. 
The  seed  photons in all sources have  temperatures kT$_0$$<$1~keV 
while the Comptonization parameter $y$, 
defined as $y$=4kT$_{\rm e}\tau^{2}/$m$_{\rm e}$c$^{2}$, is in the range
$\sim$1--7, except for the very low value of $\sim$0.03 for the source in
NGC\thinspace6441, that might be not in the right regime, since $y$$\ll$1 and $\tau$$<$1.
The trend of {\em y} with the luminosity 
is shown in Fig.~\ref{fig:lum_coptt}, 
suggesting that at least for some sources 
a correlation  exists.
The dependence of the plasma optical depths $\tau$ and of the 
electron temperatures kT$_{\rm e}$
from the total  luminosities 
are shown in Figs.~\ref{fig:tau_lum} and~\ref{fig:kte_lum}.
 
Following in't Zand et al. (\cite{i:99}), an equivalent spherical 
radius R$_{\rm W}$ of the emission area
of the seed photons can be defined as 
R$_{\rm W}=3\times10^4 d \sqrt{\frac{f_{\rm comptt}}{1+y}}/(kT_{0})^2$~km,
where d is the distance (in kpc), f$_{\rm comptt}$ is the comptonized flux 
(in erg~cm$^{-2}$~s$^{-1}$, corrected for the 
interstellar absorption, 0.1--100 keV band) and kT$_{0}$ is the seed temperature in keV.
R$_{\rm W}$ ranges from few km 
up to $\sim$1300~km  (NGC\thinspace7078) and seems to correlate 
with the total luminosity (see Fig.~\ref{fig:rw_lum}), 
as R$_{\rm W}$ $\propto$ L$^{0.5}$ .

This might suggest that the seed photons originate  
from the boundary layer 
(the radial extent and height of which 
become larger with increasing accretion rates; 
e.g. Popham \& Sunyaev \cite{ps:00}) 
rather than  from the accretion disk.  
The comparison between R$_{\rm W}$ and the lower limit R$_{\rm in}$ to the  
inner radius  of 
the accretion disk is shown in Fig.~\ref{fig:rw_rin2}.
 
In this respect, the comparison between the temperatures  kT$_{\rm 0}$ 
of the seed photons and   
the temperature kT$_{\rm in}$ is interesting 
(see Fig.~\ref{fig:ktin_kt0}): among the globular cluster
sources, only the ultra-compact   binaries 
(located in NGC\thinspace1851, NGC\thinspace6624 and NGC\thinspace6712)
display temperatures  kT$_{\rm 0}$ consistent with   
the temperature of the inner regions
of the accretion disk.

\begin{figure}
\hbox{\hspace{-0.0cm} 
\includegraphics[width=8.5cm,angle=0]{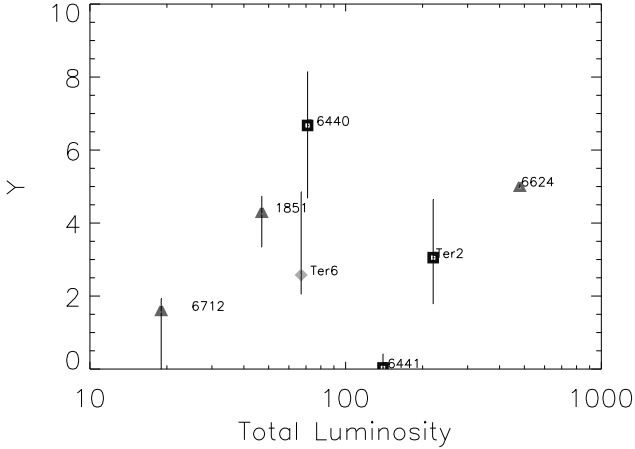}}
  \caption[]{The Comptonization parameter {\em y} versus the total luminosity in the 0.1--100 keV
  band (in units of $10^{35}$~erg~s$^{-1}$)   
}
  \label{fig:lum_coptt}
\end{figure}

\begin{figure}
\hbox{\hspace{-0.0cm} 
\includegraphics[width=8.5cm,angle=0]{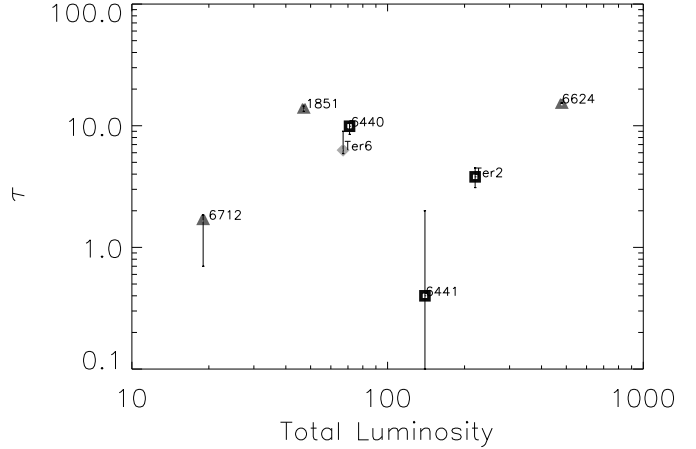}}
  \caption[]{The optical depth $\tau$ of the Comptonizing plasma 
  versus total luminosity  (0.1--100 keV, 
  in units of $10^{35}$~erg~s$^{-1}$)  
}
  \label{fig:tau_lum}
\end{figure}

\begin{figure}
\hbox{\hspace{-0.0cm} 
\includegraphics[width=8.5cm,angle=0]{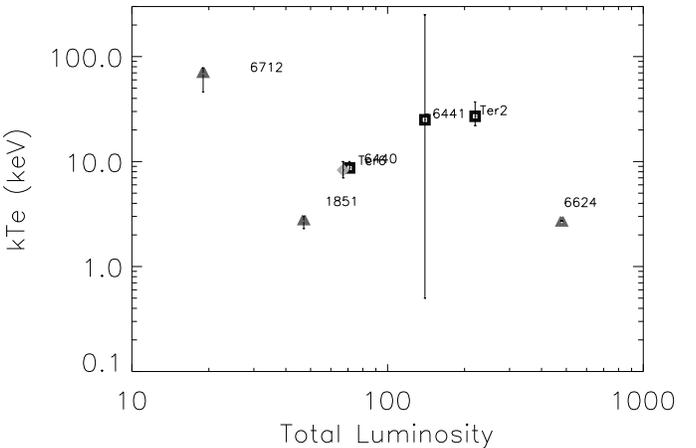}}
  \caption[]{The electron temperature  kT$_{\rm e}$  versus the total luminosity  (0.1--100 keV, 
  in units of $10^{35}$~erg~s$^{-1}$)  
}
  \label{fig:kte_lum}
\end{figure}

\begin{figure}
\hbox{\hspace{0.0cm}  
\includegraphics[width=8.5cm,angle=0]{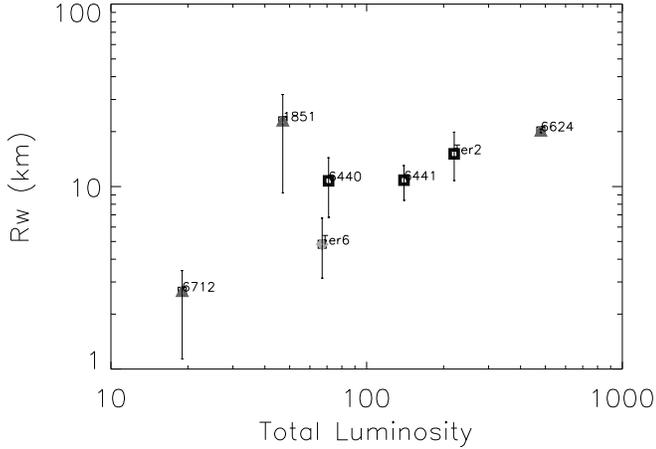}}
  \caption[]{The radius R$_{\rm W}$ of the emission area
of the seed photons  versus the total luminosity 
in the 0.1--100 keV band (in units of $10^{35}$~erg~s$^{-1}$). 
Sources with larger luminosities (i.e. accretion
rates) display larger radii of the seed photons regions  
}
  \label{fig:rw_lum}
\end{figure}

\begin{figure}
\hbox{\hspace{-0.2cm}  
\includegraphics[width=9.3cm,angle=0]{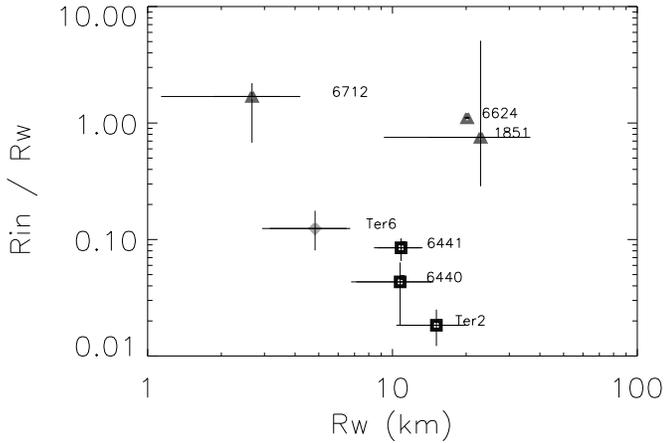}}
  \caption[]{Plot of the
ratio between the lower limit R$_{\rm in}$ of the inner radius of the accetion
disk and the radius  R$_{\rm W}$ of the emission area of the seed photons, versus
R$_{\rm W}$  
}
  \label{fig:rw_rin2}
\end{figure}

\begin{figure}
\hbox{\hspace{0.0cm}  
\includegraphics[width=8.5cm,angle=0]{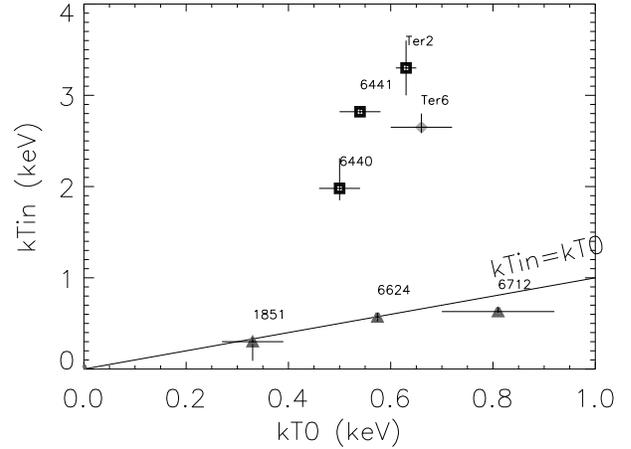}}
  \caption[]{Comparison between the  temperature of the seed photons, as estimated from the
  Comptonization model, and the temperature kT$_{\rm in}$ at 
   the inner radius of the accretion disk. The three ultra-compact   binaries, 
   indicated by the triangles, have all a  temperature of the seed photons consistent with
   originating from the accretion disk, while all other sources display a much higher inner disk 
   temperature 
}
  \label{fig:ktin_kt0}
\end{figure}

\subsection{Connection with the globular clusters metallicity}
\label{subsect:metal}

The galactic globular clusters hosting  bright LMXBs   are found to be both 
denser (except NGC~6712) and more metal-rich (except NGC~7078) 
(see Fig.~\ref{fig:c_metal}),
indicative of
the fact that both the high density stellar environment 
and the metallicity could play a
role in the formation mechanism of LMXBs, thought to be 
via tidal captures or ``two plus one" stellar 
encounters (see Bellazzini et al. \cite{b:95}).

\begin{figure}
\hbox{\hspace{0.0cm}  
\includegraphics[width=8.5cm,angle=0]{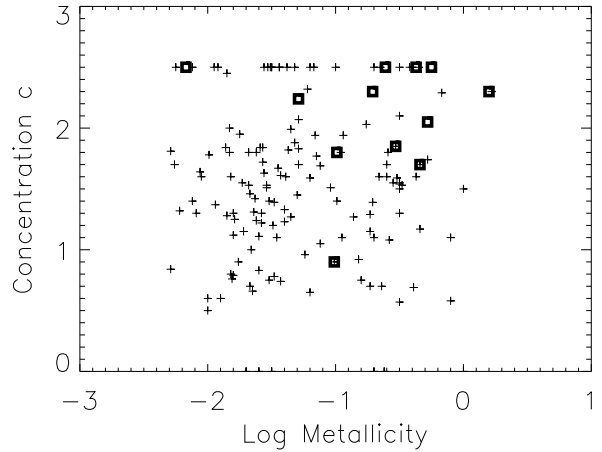}}
  \caption[]{Globular clusters 
concentrations ``c" 
(c=Log(cluster tidal radius)/Log(core radius))
versus their   metallicities ([Fe/H]). 
The  12 galactic globular clusters hosting  bright LMXBs   
(squares) are located in the 
upper right region, being denser and more metal-rich 
than the other galactic globular
clusters (marked by crosses) (see Bellazzini et al.  \cite{b:95}) 
}
  \label{fig:c_metal}
\end{figure}

Besides the formation  mechanism, it is interesting to investigate if
the metallicity of the accreting matter (assumed to be the same of the
hosting globular cluster) plays a role in the LMXBs   spectral properties.
  
In Figs.~13--19 several parameters derived from the spectral analysis are
plotted versus the metallicity ([Fe/H]) of the hosting globular cluster.
  
The positive correlation between disk temperature  kT$_{{\rm in}}$ and 
the metallicity displayed in Fig.~\ref{fig:ktinmetal} is opposite to 
the results of Irwin \&  Bregman (\cite{ib:99}), 
who studied  in the ROSAT energy band the 
dependence of the   X--ray properties  of  LMXBs  on
the metallicity of their environment, and found that 
the spectra of LMXBs   become softer as the metallicity
of their environment increases.
However, we note that our correlation has a quite large 
probability of 
11\% to be obtained  by chance; moreover, it might be spurious also
for the reason that the three ultra-compact   binaries, where
  the accreting matter should be essentially helium, are not expected  
  to show a trend with the metallicity of the hosting globular cluster. 
Note that the same effect may be present  when considering the 
 apparent correlation 
between the total X--ray luminosity and the metallicity
(see Fig.~\ref{fig:lummetal}, probability of chance correlation
of 20\%). Moreover, no  correlation 
is evident within the much more numerous 
X--ray sources residing in the M31 globular clusters 
(Verbunt et al. \cite{v:84}), in the same range of metallicity.

An apparent anticorrelation of the temperature of the seed photons kT$_0$ with
the metallicity is present, for sources in globular clusters more metallic than 0.1 solar
(see Fig.~\ref{fig:kt0metal}).
For  the electron temperature kT$_{\rm e}$ and the
optical depth of the comptonizing plasma $\tau$  there are
 no obvious correlations with the metallicity
of the accreting matter.

\begin{figure}
\hbox{\hspace{0.0cm}  
\includegraphics[width=8.5cm,angle=0]{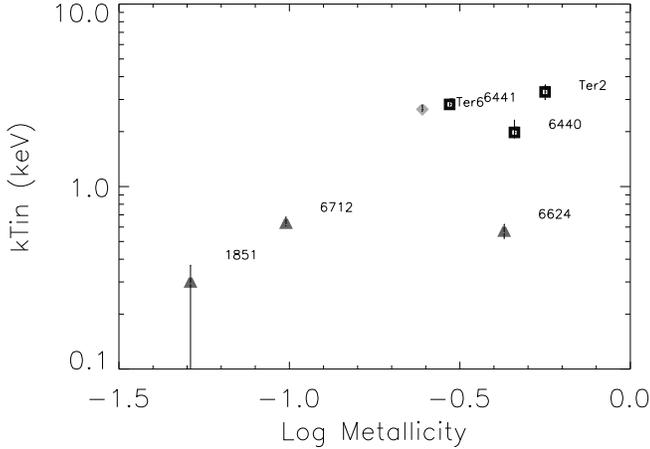}}
  \caption[]{Variation of the disk-blackbody temperature kT$_{{\rm in}}$ with the
  globular cluster metallicity ([Fe/H]). 
  The accretion disk temperatures become higher as the  metallicity increases. 
  However this correlation may be spurious, since the three ultra-compact   binaries, where
  the accreting matter is essentially helium, are not expected  
  to show a trend with the metallicity 
}
  \label{fig:ktinmetal}
\end{figure}

\begin{figure}
\hbox{\hspace{0.0cm}  
\includegraphics[width=8.5cm,angle=0]{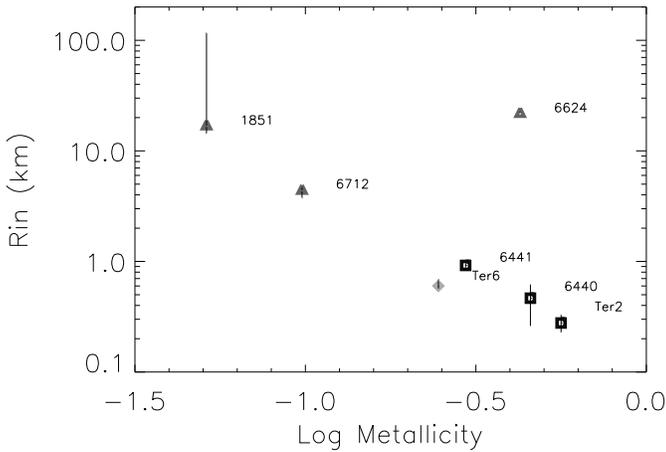}}
 \caption[]{Variation of the lower limit to the disk-blackbody inner radius  with the
  globular cluster metallicity ([Fe/H]) 
}
  \label{fig:rinmetal}
\end{figure}

\begin{figure}
\hbox{\hspace{0.0cm}  
\includegraphics[width=8.5cm,angle=0]{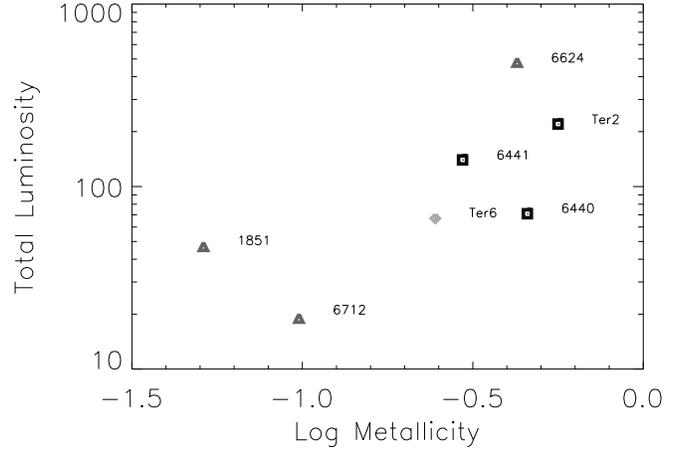}}  
\caption[]{Total luminosity 
(0.1--100 keV, in units of $10^{35}$~erg~s$^{-1}$) 
versus globular clusters metallicity ([Fe/H]).
  A possible correlation exists, 
  with higher luminosities as the metallicity increases. 
However the same caution as for the 
  possible correlation of kT$_{\rm in}$ with [Fe/H] 
should be reminded (see Fig.~\ref{fig:ktinmetal}) 
}
  \label{fig:lummetal}
\end{figure}

\begin{figure}
\hbox{\hspace{0.0cm}  
\includegraphics[width=8.5cm,angle=0]{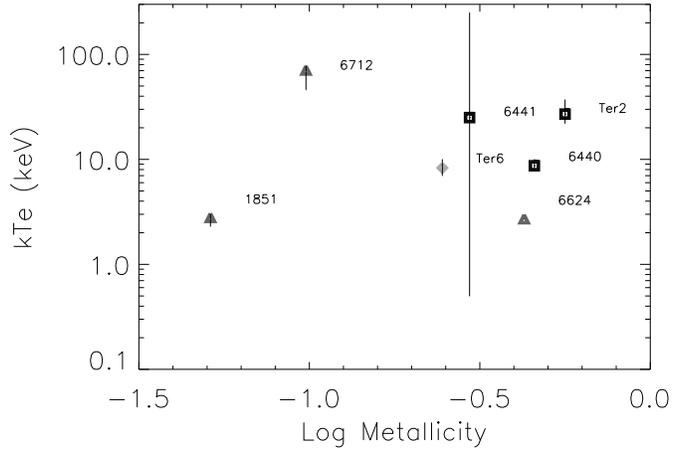}}
\caption[]{The electron temperature  kT$_{\rm e}$  versus
		the globular cluster metallicity ([Fe/H]) 
}
  \label{fig:ktemetal}
\end{figure}

\begin{figure}
\hbox{\hspace{0.0cm}  
\includegraphics[width=8.5cm,angle=0]{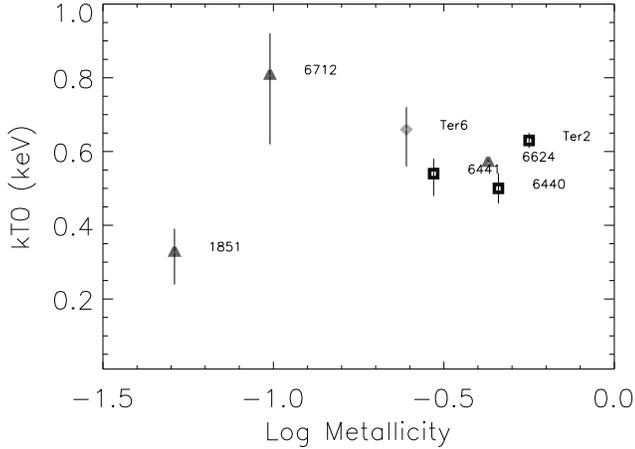}}
\caption[]{The  temperature  kT$_{0}$  of the seed photons in the Comptonization model, 
                versus
		the globular cluster metallicity ([Fe/H]). An  anticorrelation exists for globular
		clusters more metallic than about 0.1 solar
}
  \label{fig:kt0metal}
\end{figure}

\begin{figure}
\hbox{\hspace{0.0cm}  
\includegraphics[width=8.5cm,angle=0]{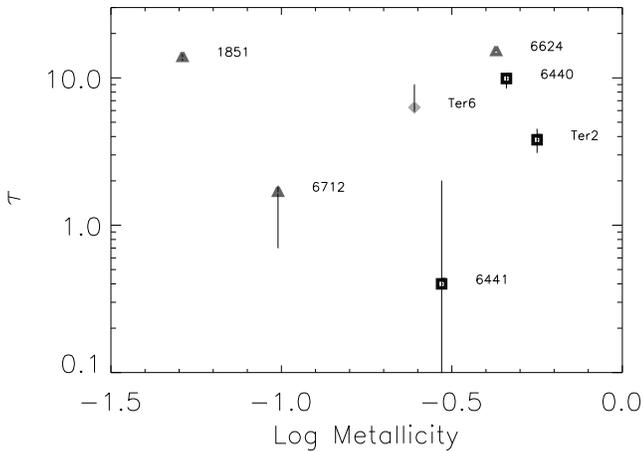}}
\caption[]{The plasma optical depth  $\tau$   
                versus
		the globular cluster metallicity ([Fe/H]) 
}
  \label{fig:taumetal}
\end{figure}

\begin{figure}
\hbox{\hspace{0.0cm}  
\includegraphics[width=8.5cm,angle=0]{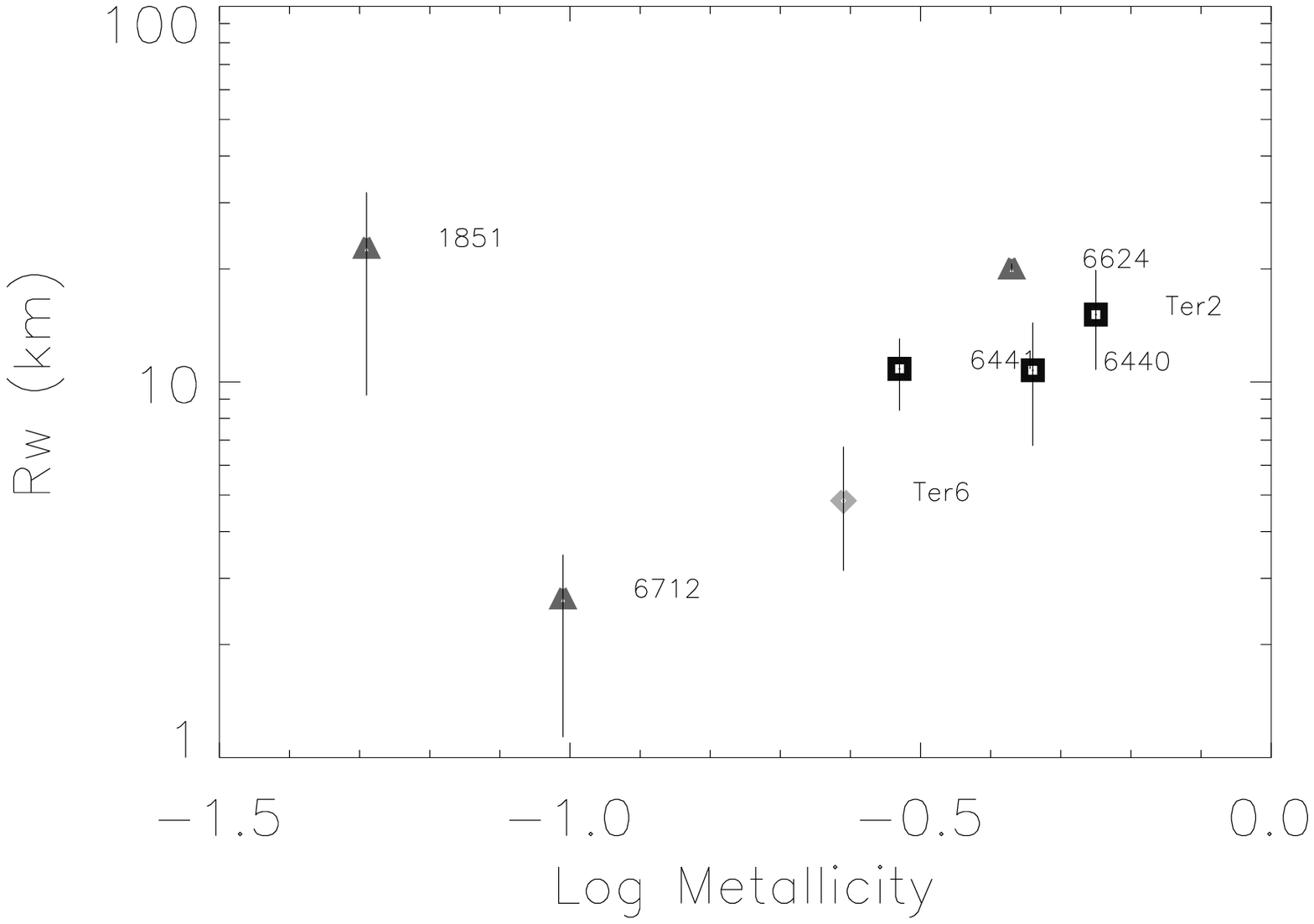}}
\caption[]{The radius R$_{\rm W}$ of the region of the seed photons   
                versus
		the globular cluster metallicity ([Fe/H]). The correlation in the metallicity range
		0.1-1 solar originates from the  
		anticorrelation between kT$_0$ and the metallicity in that
		range 
 }
  \label{fig:rwmetal}
\end{figure}

\subsection{Iron line emission}
\label{subsect:iron}

All the sources analysed here do not display significant iron line emission, except
for the source in NGC~6624, where a very weak narrow line is detected. 
The addition of a narrow K$_{\rm \alpha}$ iron line to the standard model,
resulted   in the upper limits to the  EW
reported in   Table~\ref{tab:iron}.

\begin{table}[h!]
\caption[]{Equivalent Width of a narrow 
iron line (energy fixed at 6.7 keV) 
added to the standard model. Upper limits are at the 95\% confidence level }
\begin{flushleft}
\begin{tabular}{lc}
\hline\noalign{\smallskip}
Cluster  &    EW (eV)  \\
\noalign{\smallskip\hrule\smallskip}
NGC\thinspace1851 & $<$11  \\
Terzan\thinspace2 &  $<$25 \\
Liller\thinspace1 &  $<$105 \\
Terzan\thinspace1&  $<$2200 \\
NGC\thinspace6440 & $<$45 \\
NGC\thinspace6441 &  $<$15 \\
Terzan\thinspace6 & $<$25 \\
NGC\thinspace6624 &  13$\pm ^{12} _{11}$  \\
NGC\thinspace6712 & $<$13 \\
NGC\thinspace7078  & $<$33 \\
\noalign{\smallskip\hrule\smallskip}
\end{tabular}
\end{flushleft}
\label{tab:iron}
\end{table}

\section{Discussion}
\label{sect:discussion}

We have performed a systematic survey of the properties of the luminous
X--ray sources located in globular clusters.
A remarkable property of this sample of LMXBs   containing 
neutron stars is their
well known distance and interstellar absorption.
This, together with the broad-band capabilities of the 
BeppoSAX Narrow Field Instruments, 
allowed us to properly model   the 0.1--100 keV spectra, to accurately 
determine  the X--ray  luminosity for these sources 
and to study their general
spectral properties  in a range of luminosity spanning more than two
order of magnitudes.

The spectral fitting gave satisfactory results with our spectral ``standard"
model, consisting of   a disk-blackbody plus a Comptonized emission.
Other two-component models, such as a cut-off   power-law 
plus a single-blackbody 
or  a single-blackbody plus Comptonization emission,
almost always resulted in  poorer quality fits.

However, although our standard  model provides a 
good fit to the spectra of 
almost all the LMXBs   studied, several problems arise with the
 physical interpretation of the
resulting spectral  parameters, especially when 
dealing with the disk-blackbody
component.

Indeed, the resulting inner disk radii are too small, 
even for extreme inclination
angles of the accretion disks, and, on the other hand, 
the inner disk temperature
are high compared with other systems in the galactic plane, 
for which  kT$_{\rm in}$$\sim$1.5 is usually found 
(e.g. Tanaka \& Lewin \cite{t:95}).
For comparison,   this same spectral model  has been applied 
to the \sax\ data of two X--ray bursters,
Ser~X--1 and GX~3+1, finding  a ${\rm r_{in}({\cos}i)^{0.5}}$ of
6.8$\pm{0.8}$~km and 2.8$\pm ^{1.4} _{1.2}$~km, respectively,
and a disk temperature, kT$_{\rm in}$, of 1.46 and 1.95 keV 
(Oosterbroek et al. \cite{o:00}).

The only globular cluster sources for which the 
multicolor disk-blackbody model 
seems to  reasonably represent  the likely
physical parameters of the disk 
are the ultra-compact binaries, located in NGC~1851,
NGC~6712 and NGC~6624.
In these three sources the disk-blackbody provides 
reasonable disk radii, and lower 
inner disk temperatures. 
Moreover, since in these three sources the 
temperature of the seed photons 
kT$_0$ and the radius R$_{\rm W}$
are in agreement with kT$_{\rm in}$ and r$_{\rm in}$ respectively, 
it is tempting 
to identify the inner disk as the region for the seed photons of 
 the Comptonization emission.
However, we note that the finding of a lower disk temperature kT$_{\rm in}$
 for the ultra-compact   binaries
could be partly due to the low interstellar absorption (as estimated
from optical observations) towards these sources,
with respect to other globular cluster sources (see Fig.~\ref{fig:ktin_nhopt}).

\begin{figure}
\hbox{\hspace{0.0cm}  
\includegraphics[width=8.5cm,angle=0]{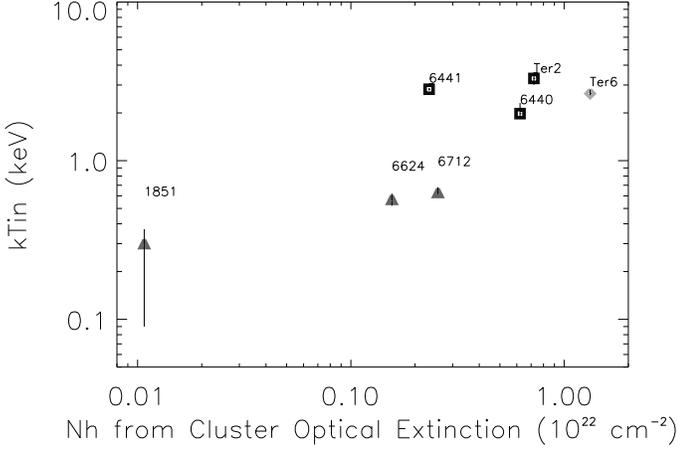}}
\caption[]{The inner disk temperature     kT$_{\rm in}$ versus the N$_{\rm H}$ estimated
from the optical absorption (see Table~1)  
}
  \label{fig:ktin_nhopt}
\end{figure}

For sources where  temperatures kT$_0$ were significantly 
different from kT$_{\rm in}$ 
(NGC~6440, NGC~6441, Terzan~2 and Terzan~6),
we tried also to fit their spectra with the ``standard" model
linking  the temperatures kT$_0$   and kT$_{\rm in}$  together. 
The main result of this   fit with respect to 
those reported in Table~\ref{tab:spec} is the values of the 
disk-blackbody
parameters:  the  temperatures kT$_0$=kT$_{\rm in}$
were in the range 0.5--0.8 keV, the ratio f  was less than 0.15 for
all the three sources
(remarkable, since previous analysis of the NGC~6441 spectrum 
resulted in a ratio f=10) and the disk-blackbody radii became 1--8.5~km.
However, this alternative fit  resulted
in   worse \chisq\ except for NGC~6440, for which an  equally good \chisq\ was
obtained.

Another problem with the standard model arises 
from the relative contribution of the
disk to the  Comptonized component. 
Indeed, the ratio f of the disk-blackbody 0.1--100 keV
luminosity to the Comptonized one, 
should be $\sim$1 (Sunyaev \& Shakura  \cite{ss:86}). 
But we find f$>$1 for the sources  NGC~7078 (the ADC source; f$\sim$2), 
Terzan~6 (f$\sim$2) up to the very 
high value of NGC~6441 (f=10). 
On the other hand, 
when a   single-blackbody is used instead of the disk-blackbody,
the ratio f$_{bb}$  
is always less than 1 (see Table~\ref{tab:spec2}).
NGC~6441 is an intriguing source: 
if a blackbody is used instead of
the disk-blackbody, the Comptonization dominates 
the emission and the high energy region
of the spectrum, whereas when a disk-blackbody model is used, 
the opposite is valid.

For the LMXBs   where the interpretation of the soft component as
contributed by the inner disk is reasonable, it is possible to estimate
the strength of the neutron star magnetic field.
The total luminosity L$_{\rm tot}$ of an accreting neutron star can 
be expressed as the sum of the luminosity liberated
in the accretion disk L$_{\rm disk}$ and the remaining luminosity dissipated
at the neutron star radius R$_{\rm ns}$, 
where L$_{\rm tot}$=GM$_{\rm ns}$$\mdot$/R$_{\rm ns}$ and 
L$_{\rm disk}$=GM$_{\rm ns}$$\mdot$/2R$_{\rm m}$ 
(e.g., Priedhorsky \cite{p:86}). 
R$_{\rm m}$ is the magnetospheric radius 
R$_{\rm m}=20\,{\rm B}_8^{4/7}\,{\rm R}_6^{10/7}\,{\rm M}_2^{1/7}\,{\rm L}_{37}^{-2/7}\,\xi$~km, where
B=B$_8\,10^8$~G is the magnetic field, R$_{\rm ns}$=R$_6$\,10 km, 
M$_{\rm ns}$=2\,M$_2$$\msun$ is the neutron star mass, 
L$_{\rm tot}=10^{37}\,$L$_{37}$~erg~s$^{-1}$ and the parameter $\xi$ accounts
for the disk geometry of the accretion and is $\xi$$\sim$0.5--1, 
(Ghosh \& Lamb \cite{gl:79}, \cite{gl:92}).
Thus, the  magnetic field strength of  a neutron star  
can be estimated from the luminosity of the disk.

In Fig.~\ref{fig:fdisktot} a grid with different neutron star 
magnetic fields, calculated assuming ${\rm R}_6$=1, ${\rm M}_2$=1 and 
$\xi$=1, is superimposed on a plot of the ratio of disk to total
luminosity, versus the total luminosity (0.1--100 keV).
The ratio of L$_{\rm disk}$ to 
L$_{\rm tot}$ is  a measure of R$_{\rm ns}$/2R$_{\rm m}$ 
(Priedhorsky \cite{p:86}):

\begin{equation}
\frac{L_{\rm disk}}{L_{\rm tot}} = 0.25 {\rm B}_8^{-4/7}\,{\rm R}_6^{-3/7}\,{\rm M}_2^{-1/7}\,{\rm L}_{37}^{2/7}\,\xi
\end{equation}
 
\noindent
Assuming a 2$\msun$ neutron star, this ratio depends only on the 
magnetic field strength and total luminosity. 
The ratio of L$_{\rm disk}$ to L$_{\rm tot}$ reaches a
maximum of 0.5 when R$_{\rm m}$ is equal to R$_{\rm ns}$; since the
sources in NGC~6441, NGC~7078 and Terzan~6 have an  unusually high
disk-blackbody emission (as estimated from the disk-blackbody model
used here), that led to ratios above the maximum allowed of 0.5, 
we will not consider them.

The magnetic field estimated in this way for the 
source X\thinspace1820--30 in NGC\thinspace6624 (2.1$\times$10$^{8}$~G) 
is in agreement with the value derived from the saturation 
of the kHz QPO frequencies (Campana \cite{ca:00}). 
In the case of the RB, we find a magnetic field ($\sim$8$\times10^{8}$~G)
consistent, within a factor of 2, with that
estimated by Masetti et al. (\cite{m:00}) using the propeller effect model.

\begin{figure}
\hbox{\hspace{0.0cm}  
\includegraphics[width=8.5cm,angle=0]{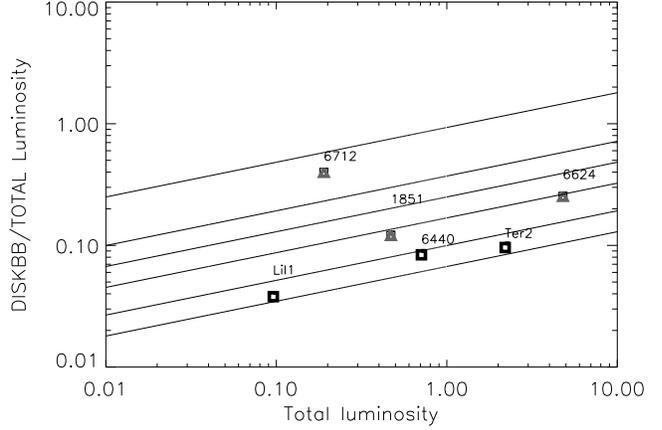}}
\caption[]{An estimate of the neutron star magnetic field strength 
depending on
the disk and total luminosities in the 0.1--100 keV, 
in units of $10^{37}$~erg~s$^{-1}$. 
Straight lines are loci of constant magnetic field 
(0.1, 0.5, 1, 2, 5 and 10$\times$10$^{8}$~G, increasing from top to bottom), 
assuming a 2$\msun$ neutron star. 
A lower neutron star mass results in a 
shift of the magnetic field grid towards the bottom of this plot. 
Only sources with f$<$1 are reported
}
\label{fig:fdisktot}
\end{figure}

\section{Conclusions}
\label{sect:conclusions}

We performed  broad--band spectroscopy of the luminous X--ray sources 
located in a number of  galactic globular clusters 
with \sax.
The  results can be  summarized as follows:

\begin{itemize}

\item All the spectra can be acceptably fit by the same two--component
model, comprising  high energy Comptonization emission 
together with 
a disk-blackbody accounting for the soft part of the spectrum, except
for the source in NGC~6441, where the disk--blackbody dominates
the high energy region.
\\

\item In the framework of this ``standard" model and 
as a result of the spectral fitting, the sources can be clearly 
divided into {\em two
groups},  
 on the basis of their disk--blackbody 
parameters: 
(1) the sources located in NGC~1851, NGC~6712 and NGC~6624,
and 
(2) all the other sources.
Indeed, the disk--blackbody model, if literally interpreted, provides 
a physically reasonable interpretation for the first group of sources,
while for the second  group the disk--blackbody spectral parameters are not physically
acceptable, 
displaying high inner disk temperatures and  extremely low innermost radii 
of the accretion disks, too low
even for edge-on binary systems.  
There is no obvious explanation for a factor of $\sim$3--10 lower  disk temperatures 
in the ultra-compact binaries with respect to all the other
systems; higher neutron star masses in ultra-compact binaries would 
contribute in reaching lower inner disk temperatures kT$_{\rm in}$, 
but cannot explain,
alone, a factor of $\sim$3--10 lower  disk temperatures  with respect to
other binary systems. Indeed, for systems with similar disk-blackbody
luminosities, the ratio of the neutron star masses 
M$_{\rm ns 1}$/M$_{\rm ns 2}$$\propto$$({\rm kT_{in2}}/{\rm kT_{in1}})^{2}$,
thus a factor of $\sim$3--10 in inner temperatures translates in a factor
of 10--100 in neutron star masses.
\\

\item For the first group of sources, 
the Comptonization emission has
seed photons temperatures kT$_{\rm 0}$ and radii R$_{\rm W}$  
consistent with their inner disk temperatures kT$_{\rm in}$
and radii for 
the accretion disks ${\rm r_{in}({\cos}i)^{0.5}}$, respectively.
This could be interpreted with  Comptonization 
sustained and fed by the inner regions of the accretion disks.
Interestingly, this first group of sources is composed of   
the ultra-compact binary systems.
If this result is indicative of a spectral signature
of the ultra-compact binaries, we suggest  that the ultra-compact
binary located in NGC~6652 
will
display the same spectral behavior and that,
among the sources here studied for which 
both the orbital period and the optical
counterpart are unknown, no other ultra-compact binary systems 
are present.
\\

\item In all other sources (group 2), the temperatures of the 
seed photons (kT$_{\rm 0}$) are significantly lower than the 
temperatures kT$_{\rm in}$. 
\\

\item  There is no strong evidence for a correlation 
of Comptonization spectral
parameters with total luminosity L (0.1--100 keV), except for
the radius R$_{\rm W}$ of the emission area
of the seed photons, for which  R$_{\rm W}$ $\propto$ L$^{0.5}.$
\\

\end{itemize}

The search for correlations of the spectral parameters 
with  the metallicity of the hosting globular clusters
led to the following results:

\begin{itemize}

\item The temperature kT$_{\rm in}$ of the inner radius increases as the
metallicity of the hosting cluster increases, and   
the opposite is true for the inner disk  radius, except for 
the source in NGC~6624. This trend could be spurious, since in the 
three ultra-compact binaries the accreting matter might be 
only helium. 
On the other hand, if this is real, it suggests that 
matter accreting in the ultra-compact binaries
NGC~1851 and NGC~6712 is probably not pure helium. 
\\

\item The total luminosity L (0.1--100 keV) seems to increase with the
metallicity, but again this trend could be spurious, since it is mainly based
on the behavior of the ultra-compact binaries.
\\

\item There is no clear evidence for 
a correlation of the Comptonization spectral
parameters with   metallicity.
\\

\item We caution that an observational bias could be present
due to the galactic absorption
in the direction of the globular cluster sources: 
lower disk--temperatures kT$_{\rm in}$ have been derived for sources 
with low interstellar absorption, and these happen to be the ultra-compact binaries.

\end{itemize}

In a literal  interpretation  of the disk-blackbody model, 
the  luminosity of the accretion disk, 
for the sources 
with a ratio f$<$1, is used to derive 
an estimate of the neutron star magnetic field.
We note that, for the two sources for which an independent estimate
of the magnetic field exists (the RB  
and X1820--303, located in NGC~6624), 
there is a good agreement with the values derived here.

On the other hand, problems exist with a physical interpretation
of the   spectral parameters
of the disk-blackbody component, since   we obtained  
resonable  values for the   radii
of the inner disks    only in the case of the ultra-compact   binaries.
From the similarities between   radii and temperatures of the inner accretion
disk and the parameters of the ``seed" photons in the Comptonization
emission, we propose that in the ultra-compact   binaries in 
NGC~1851,
NGC~6712 and NGC~6624 the Componization is fed by the inner regions
of the accretion disk, while this is not the case for all other
sources (in NGC~6440, NGC~6441, Terzan~2 and Terzan~6).
It is unclear why the disk-blackbody model gives physically reasonable
results only in  the case of the ultra-compact binaries.
A possibility may be related to the mass ratio, q, 
which is similar 
in the ultracompact binaries and binary systems containing black holes. 
Indeed, physically reasonable parameters
are usually obtained from disk-blackbody fits to
the soft emission from black hole X--ray binaries 
(see, e.g., Tanaka \& Lewin \cite{t:95}).

\begin{acknowledgements}
The \sax\ satellite is a joint Italian-Dutch programme. 
We thank the staffs of the \sax\ Science Data and
Operations Control Centers for help with these observations. 
L.~Sidoli acknowledges an ESA Fellowship. 
\end{acknowledgements}



\begin{thebibliography}{}

\bibitem[1989]{a:89}
Anders E., Grevesse N., 1989, Geochimica et Cosmochimica Acta 53, 197 
 

\bibitem[1997]{ba:97}
Barbuy B., Ortolani S., Bica E., 1997, A\&AS 122, 483

 

\bibitem[1995]{b:95}
Bellazzini M., Pasquali A., Federici L., et al., 1995, ApJ 439, 687
 
\bibitem[1997]{b:97}
Boella G., Chiappetti L., Conti G., et al., 1997, A\&AS 122, 327
 

\bibitem[2000]{ca:00}
Campana S., 2000, ApJ 534, L79

\bibitem[2000]{c:00}
Carretta E., Gratton R.G., Clementini G., Fusi Pecci F., 2000, ApJ 533, 215
 

\bibitem[1993]{d:93}
Djorgovski S., 1993, ASP Conf. Ser. 50, 373
 
\bibitem[1999]{f:99} 
Fiore F., Guainazzi M., Grandi P., 1999, Technical Report version 1.2 (BeppoSAX Scientific Data Center)
  
\bibitem[1987]{f:87}
Frank J., King A.R., Lasota J.P., 1987, A\&A 178, 137

\bibitem[1995]{f:95}
Frogal J.A., Kuchinski L.E., Tiede G.P., 1995, AJ 109, 3, 1154

\bibitem[1997]{f:97}
Frontera F., Costa E., Dal Fiume D., et al., 1997, A\&AS 122, 371
 

\bibitem[1979]{gl:79}
Ghosh P., Lamb F.K., 1979, ApJ 234, 296

\bibitem[1992]{gl:92}
Ghosh P., Lamb F.K., 1992, in  van  den Heuvel E.P.J.,  Rappaport S.A. (eds.) 
``X--ray Binaries and Recycled Pulsars'',  (Dordrecht: Kluwer), p.~487
 

\bibitem[1998]{g:98}
Guainazzi M., Parmar A.N., Segreto A., et al., 1998, A\&A 339, 802

\bibitem[1999]{g:99}
Guainazzi M., Parmar A.N., Oosterbroek T., 1999, A\&A 349, 819
 

\bibitem[1985]{h:85}
Hertz P., Wood K.S., 1985, ApJ 290, 171    
 

\bibitem[1995]{h:95}
Hua X.-M., Titarchuk L., 1995, ApJ 449, 188
  
\bibitem[1999]{i:99}
In 't Zand J.J.M., Verbunt F., Strohmayer T.E., et al., 1999,
A\&A 345, 100

\bibitem[2000]{i:00}
In 't Zand J.J.M., Bazzano A., Cocchi M., et al., 2000, A\&A 355, 145
 
\bibitem[1999]{ib:99}
Irwin J.A., Bregman J.N., 1999, ApJ 510, L24 
 
\bibitem[1986]{m:86}
Makishima K., Maejima Y., Mitsuda K., et al., 1986, ApJ 285, 712           

 
\bibitem[2000]{mkm:00}
Makishima K., Kubota A., Mizuno T., 2000, ApJ 535, 632

\bibitem[1997]{m:97}
Manzo G., Guarrusso S., Santangelo A., et al., 1997, A\&AS 122, 341
 
\bibitem[2000]{m:00}
Masetti N., Frontera F., Stella L., et al., 2000,  A\&A in press
 
\bibitem[2000]{mfr:99}
Merloni A., Fabian A.C., Ross R.R., 2000, MNRAS 313, 193                  
    
\bibitem[1984]{m:84}
Mitsuda K., Inoue H., Koyama K., et al., 1984, PASJ 36, 741            
 
\bibitem[1983]{m:83}
Morrison D., McCammon D., 1983, ApJ 270, 119
 

\bibitem[1994]{o:94}
Ortolani S., Bica E., Barbuy B.,  1994, A\&A 286, 444

\bibitem[1999]{o:99}
Ortolani S., Barbuy B., Bica E., et al., 1999, A\&A 350, 840


\bibitem[2000]{o:00}
Oosterbroek T., Barret D., Guainazzi M., Ford E.C., 2000, preprint (astro-ph/0010609), 
A\&A in press


\bibitem[1989]{p:89}
Parmar A.N., Stella L., Giommi P., 1989, A\&A 222, 96  

\bibitem[1997]{p:97} 
Parmar A.N., Martin D.D.E., Bavdaz M., et al., 1997, A\&AS 122, 309

\bibitem[1999]{p:99} 
Parmar A.N., Oosterbroek T., Guainazzi M., et al., 1999, A\&A 351, 225
 
\bibitem[2000]{ps:00}
Popham R., Sunyaev R., 2000, preprint (astro-ph/0004017), ApJ submitted
 
\bibitem[1995]{ps:95}
Predehl P., Schmitt J.H.M.M., 1995, A\&A 293, 889

\bibitem[1986]{p:86}
Priedhorsky W., 1986, ApJ 306, L97
 
\bibitem[2000]{s:00}
Sidoli L., Parmar A.N., Oosterbroek T., 2000, A\&A 360, 520

 
\bibitem[2000b]{s:00b}
Sidoli L., Parmar A.N., Oosterbroek T., et al., 2000b,  
in Giacconi R., Stella L., Serio S. (eds.),
Proc. of  ``X--ray Astronomy 2000", Palermo, Sept. 2000,  
 ASP Conf. Ser. in press  

 
 
\bibitem[1986]{ss:86}
Sunyaev R., Shakura N.I., 1986, Sov. Astron. Lett. 12, 117 
 
\bibitem[1995]{t:95}
Tanaka Y., Lewin W.H.G., 1995, in Lewin W.H.G., van  Paradijs J., 
van  den Heuvel E.P.J. (eds.) ``X--ray Binaries", Cambridge University Press, Cambridge, p.~126 

\bibitem[1994]{ti:94}
Titarchuk L., 1994, ApJ 434 570
 
\bibitem[1995]{tl:95}
Titarchuk L., Lyubarskij Y., 1995, ApJ 450, 876
 
\bibitem[1984]{v:84}
Verbunt F., van  Paradijs J., Elson R., 1984, MNRAS 210, 899

\bibitem[1995]{v:95}
Verbunt F., Bunk W., Hasinger G., Johnston H., 1995, A\&A 300, 732
 
\bibitem[1988]{w:88}
White N.E., Stella L., Parmar A.N., 1988,  ApJ 324, 363             
           
\end{thebibliography}
\end{document}